# Nd induced Mn spin-reorientation transition in NdMnAsO

A. Marcinkova,[1,2] T.C. Hansen,[3] C. Curfs,[4] S. Margadonna,[1,2] and J.W.G. Bos[5*]

1. School of Chemistry, University of Edinburgh, Edinburgh, EH9 3JJ
2. Centre for Science at Extreme Conditions, University of Edinburgh, King's Buildings, Mayfield Road, Edinburgh EH9 3JZ
3. Institut Laue Langevin, 38042 Grenoble, France
4. European Synchrotron Radiation Facility, 39043 Grenoble, France
5. Department of Chemistry - EPS, Heriot-Watt University, Edinburgh, EH14 4AS

* j.w.g.bos@hw.ac.uk

A combination of synchrotron X-ray, neutron powder diffraction, magnetization, heat capacity and electrical resistivity measurements reveals that NdMnAsO is an antiferromagnetic semiconductor with large Neel temperature ($T_N$ = 359(2) K). At room temperature the magnetic propagation vector **k** = 0 and the Mn moments are directed along the crystallographic *c*-axis ($m_{Mn}$ = 2.41(6) $\mu_B$). Upon cooling a spin reorientation (SR) transition of the Mn moments into the *ab*-plane occurs ($T_{SR}$ = 23 K). This coincides with the long range ordering of the Nd moments, which are restricted to the basal plane. The magnetic propagation vector remains **k** = 0. At base temperature (1.6 K) the fitted moments are $m_{ab,Mn}$ = 3.72(1) $\mu_B$ and $m_{ab,Nd}$ = 1.94(1) $\mu_B$. The electrical resistivity is characterized by a broad maximum at 250 K, below which it has a metallic temperature dependence but semiconducting magnitude ($\rho_{250K}$ = 50 $\Omega$ cm, residual resistivity ratio = 2), and a slight upturn at the SR transition.

PACS: 75.25.-j, 75.30.Kz, 75.50.Ee

**Introduction**

The discovery[1] of high-temperature superconductivity in LaFeAsO$_{1-x}$F$_x$ has generated enormous interest and led to the rapid exploration of materials with similar structures. High-T$_c$ superconductivity has since been found in 1111-type RFeAsO and AeFeAsF (R = rare-earth, Ae

= alkaline-earth), 122-type AeFe$_2$As$_2$, 111-type A$_x$FeAs (A = Li, Na) and 11-type FeSe$_{1-d}$ based materials that all have square planar Fe layers with tetrahedral As (Se) coordination.[2,3] The 1111-type materials have the highest superconducting temperatures (up to 55 K).[4,5] It is now well established that the superconductivity in these materials occurs at the expense of an itinerant antiferromagnetic (AF) spin density wave,[6-8] and is induced either by chemical doping or application of hydrostatic pressure. It is therefore of much interest to study the interplay between composition, structure, magnetism and superconductivity in these but also in other non-iron based materials. For instance, the reported Ni analogues are conventional superconductors with < 4 K transition temperatures without any evidence for the importance of magnetism in the pairing mechanism.[9-11] The isostructural Co materials are ferromagnetic (FM) metals or on the boundary of ferromagnetism.[12-15] The Mn analogues in contrast are not metallic but semiconducting and are antiferromagnetic.[16-19] This pronounced difference has been attributed to a much stronger hybridization between Mn 3d and pnictogen p states, which leads to an electronic structure that is not derived from the Fe-Ni materials.[18] The best characterized example is BaMn$_2$As$_2$, which has a G-type magnetic structure with moments aligned along the $c$-axis [3.88(4) $\mu_B$ at 10 K] and a Neel temperature of 625(1) K.[19] Heat capacity and electrical resistivity reveal a Sommerfeld coefficient close to zero, a metallic temperature dependence between 100-400 K and a room temperature (RT) resistivity of 165 m$\Omega$ cm.[17,18] Among the 1111-type materials, LaMnPO has been characterized in some detail.[16] It has a similar magnetic structure to BaMn$_2$As$_2$ but with a parallel alignment of the nearest neighbor (NN) Mn moments along the $c$-axis. Resistivity measurements on polycrystalline samples revealed Arrhenius type semiconducting behavior between 200-300 K with a RT value of 2 k$\Omega$ cm.

The 1111-type materials offer the possibility to investigate the interplay between rare-earth (R) and transition metal (TM) magnetism in this important class of materials. So far most of the experimental evidence points to a rather weak interaction between R and TM sublattices. For instance, the increase in T$_c$ on going from La to the later R in the RFeAsO based superconductors has been attributed to a structural effect rather than a strong 4f-3d hybridization.[20-22] One noteworthy exception is CeFePO, where the absence of low temperature superconductivity is related to a strong hybridization resulting in heavy Fermion behaviour.[23] In case of the Fe and Co RTMAsO materials, the most common effect of magnetic R ordering is a change in sign of the magnetic coupling between adjacent TM planes, while the in-plane ordering remains intact.[3, 14, 15, 20] For example for NdCoAsO, the coupling between adjacent Co planes changes from FM to AF upon ordering of the Nd-moments without affecting the in-plane order.

Here, we report on the crystal and magnetic structure, and physical properties of NdMnAsO using synchrotron X-ray, neutron powder diffraction, magnetization and electrical resistance measurements. Analysis of the neutron diffraction data reveals a spin-reorientation (SR) transition of the Mn moments from parallel to the *c*-axis into the basal plane upon ordering of the Nd moments. This demonstrates that a magnetic R sublattice can change the orientation of the transition metal moments, and is not limited to altering the spin-orientation between unchanged planes.

**Experimental**

Polycrystalline NdMnAsO (2 g) was prepared using solid state chemistry methods, and used for all measurements. A stoichiometric mixture of NdAs and MnO (Sigma Aldrich, 99.999%) was homogenized using mortar and pestle, pressed into pellets and heated for 24 hours at 1100 °C inside an evacuated quartz tube. NdAs was prepared from a 1:1 ratio of Nd (Sigma Aldrich, 99.99%) and As (Sigma Aldrich, 99.9%), which was heated for 2 hr at 500 °C in an evacuated quartz tube, immediately followed by 16 hrs at 900 °C. Initial phase analysis was done using laboratory powder X-ray diffraction on a Bruker D8 AXS diffractometer with a Cu K$_{\alpha1}$ radiation source. High-resolution synchrotron X-ray powder diffraction measurements were done on the ID31 beam line at the European Synchrotron Radiation Facility in Grenoble, France. The X-ray wavelength used was 0.3998 Å, the sample was contained in a 0.5 mm diameter borosilicate glass capillary, and data were collected between 4 - 250 K. The real and anomalous X-ray dispersion coefficients were calculated using the program FPrime and a linear absorption correction μR = 1.5 was used.[24] The neutron powder diffraction experiments were done on the D20 beam line at the Institute Laue Langevin in Grenoble, France.[25] The instrument was used in the high-flux setting with λ = 2.41 Å. The sample was contained in a 6 mm diameter cylindrical vanadium can. Between 1.6-30 K, 30 minute datasets were collected at fixed temperature. From 30-300 K, 5 minute datasets were collected on a ramp of 1.25 K/ minute. A linear absorption correction μR = 0.3 was applied. Rietveld analysis of the powder diffraction data was done using the GSAS/EXPGUI suite of programs.[26, 27] The zero field cooled magnetic susceptibility was measured using a Quantum Design Magnetic Property Measurement System (MPMS). The temperature and field dependences of the electrical resistivity were measured using the resistance option of a Quantum Design Physical Property Measurement System (PPMS). Heat capacity data were collected using the heat capacity option of the PPMS. A sintered platelet of about 9 mg was used, the sample coupling constant was always larger than 98%.

**Results**

Rietveld analysis of the synchrotron X-ray powder diffraction data confirms the tetragonal ZrCuSiAs structure reported by Nientiedt et al.[28] down to 4 K. The fit to the 4 K dataset is given in Fig. 1, and a summary of the fitted parameters and selected bond distances and angles at 4, 100 and 250 K are given in Table 1. Upon cooling the lattice contracts gradually along the *a*- (0.1%) and *c*-axis (0.3%) and no large anomalies are observed (Fig. 2). The c/a-ratio does evidence a change in trend at $T_{SR}$ signaling a weak coupling between the lattice and magnetic order. The Mn-As bond distance (2.5468(4) Å) and twofold As-Mn-As tetrahedral angle (111.59(2)°) do not change significantly with temperature. The "anomalous" nature of the Mn material is illustrated by its unit cell volume (146 Å$^3$), which is much larger than expected based on a simple extrapolation of the values for NdFeAsO (135 Å$^3$) and NdCoAsO (132 Å$^3$).[14, 29] Refinement of the atomic site occupancies confirmed the nominal composition.

The temperature dependence of the magnetic susceptibility ($\chi$) evidences local moment magnetism (Fig. 3) with two transitions visible at 23 and 4 K. From the neutron powder diffraction experiment (below) it is known that $T_{N, Mn}$ = 359(2) K, while the transition at 23 K corresponds to the long range AF ordering of the Nd moments, and a concomitant SR of the Mn moments. The presence of Nd and Mn moments makes interpretation of the $\chi(T)$ data difficult. In particular, the data in Fig. 3 cannot be modeled using a single Curie-Weiss term, $\chi(T) = C/(T-\theta_W)$, unless $\chi_{Mn}(T)$ is negligible. To obtain an indication of the relative importance of the Nd and Mn contributions, $1/\chi$ was fitted to a single Curie-Weiss term between 125-300 K. This yields C = 2.39(2) emu K$^{-1}$mol$^{-1}$ and $\theta_W$ = -43.1(8) K. The fitted Curie constant is larger than expected for $^4I_{9/2}$ Nd$^{3+}$ [C = 1.6 emu K$^{-1}$mol$^{-1}$], which suggests that both Nd and Mn contribute to $\chi(T)$ in this temperature interval. This is supported by the neutron diffraction results which reveal that the Mn moments are not saturated down to 50 K. The transition at 4 K corresponds to the saturation of the ordered Nd moments, and cannot be attributed to magnetic impurities (Table I) as Nd$_2$O$_3$ does not order down to at least 1.8 K, and MnAs is FM with a Curie temperature above RT. The linear field dependence of the magnetization (inset to Fig. 3) is in agreement with a complete AF (2, 15 K) or AF plus paramagnetic ground state (30 K).

The temperature and field dependences of the electrical resistivity are summarized in Fig. 4. The magnitude of the resistivity falls between 25 and 50 $\Omega$ cm, which places NdMnAsO in the semiconducting regime. The temperature dependence is characterized by a broad hump at 250 K, above which semiconducting behavior is found, while below a metallic temperature dependence is evident. At $T_{SR}$ the resistivity shows a moderate upturn, which may be due to increased spin scattering. This type of temperature dependence is typical of a degenerate semiconductor and was also observed in BaMn$_2$As$_2$ single crystals.[17] However, in the current case, the magnitude of the resistivity is 1-2 orders of magnitude larger, which may be related the presence of insulating

Nd$_2$O$_2$ blocks and to the polycrystalline nature of the samples. Measurements on single crystals are needed to probe the intrinsic transport properties of this material. The magnetoresistance (MR) was obtained from the symmetric part of R(H) curves measured between -9 T ≤ μ$_0$H ≤ 9 T. This was done to eliminate spurious Hall effect contributions due to misalignment of the contact electrodes. At low temperatures (< 30 K), the field dependence shows a rapid drop followed by a saturation at the 1-2 percent level. At higher temperatures, the field dependence is linear with a similar MR of 2% in 9 T observed at 250 K.

The temperature dependence of the heat capacity (C) in applied magnetic fields of 0 and 9 Tesla is given in Fig. 5. This confirms the presence of a phase transition at 23 K, while there is no evidence for a transition at 4 K. In zero field, the transition appears second order with a clear maximum at 23 K. Application of a magnetic field broadens the transition, and moves the maximum to lower temperatures, in agreement with the observed antiferromagnetic ordering. We note that even in zero applied magnetic field the transition appears somewhat broad, possibly signaling two 2$^{nd}$ order phase transitions, which is consistent with the neutron data that reveals a Mn-spin-reorientation occurring in a narrow temperature interval. A linear fit to C/T versus T$^2$ at low temperatures (inset to Fig. 5) yields a Sommerfeld coefficient close to zero [γ = 1.2(7) J mol$^{-1}$ K$^{-2}$], which is in agreement with the observed semiconducting behavior.

Comparison of the X-ray and neutron data revealed magnetic contributions to the nuclear (102) and (103) reflections, and completely magnetic (100) and (101) reflections in the neutron diffraction patterns from 300-25 K (Fig 5a). Below 25 K, a magnetic contribution to the (002) reflection develops, and there is a clear change in the relative intensities of the other magnetic reflections (Fig. 5b). In all cases, the magnetic reflections are indexed on the nuclear cell and the magnetic propagation vector **k** = 0 at all measured temperatures. Magnetic representational analysis was used to derive the symmetry allowed magnetic structures. These calculations were done using version 2K of the SARAh representational analysis program,[30] and confirm that all symmetry elements of the P4/nmm space group leave the magnetic propagation vector (**k** = 0) invariant, and thus constitute the small group $G_\mathbf{k}$. The decomposition of the magnetic representation ($\Gamma_{Mag}$) into the irreducible representations (IRs) of $G_\mathbf{k}$ is $\Gamma_3^1 + \Gamma_6^1 + \Gamma_9^2 + \Gamma_{10}^2$ for the Mn sites and $\Gamma_2^1 + \Gamma_3^1 + \Gamma_9^2 + \Gamma_{10}^2$ for the Nd sites. The representations used are after Kovalev,[31] and the character table can be found in Ref. 14. The symmetry allowed basis vectors are given in Table 2. Inspection of this table reveals that there are two FM ($\Gamma_3^1$ and $\Gamma_9^2$) and three AF solutions ($\Gamma_2^1, \Gamma_6^1$ and $\Gamma_{10}^2$). The two FM models are not in agreement with the observed data and can therefore be discarded. The AF models differ in the easy magnetization direction, which is either along the $c$-axis or in the basal plane. We note that according to Landau theory, only a single IR becomes critical in a 2$^{nd}$ order phase transition, and as a starting point only models corresponding to a single IR were tested. Between 300- 25 K, the magnetic intensities were best fitted using the $\Gamma_6^1$ model with the Mn moments aligned along the $c$-direction. (The $R_{wp}$ values for the tested models are given in Table II). The Rietveld fit and a graphic representation of the model are given in Fig 6a. Trial refinements using a "forbidden" linear combination of the $\Gamma_6^1$ and $\Gamma_2^1$ models did not fit the data well ($R_{wp}$ = 7.7%), and there is no evidence from our data for ordering of the Nd moments above 25 K. For T < 20 K, the magnetic intensities were fitted well using the $\Gamma_{10}^2$ model with ordered Mn and Nd moments (Table II), which are both constrained to the basal plane (Fig. 6b). The ordered moments at 1.6 K are 3.72(1) $\mu_B$ for Mn and 1.941(1) $\mu_B$ for Nd. Rietveld refinements using models with an out-of-plane component on the Nd and/or the Mn sublattice did not result in an improvement, and were generally unstable. Finally, a small intermediate temperature interval (20 ≤ T < 25 K) was identified in which the magnetic intensities were fitted using several trial models. At 21.7(2) K, the $\Gamma_6^1$ and the $\Gamma_{10}^2$ models are indiscernible ($R_{wp}$ = 5.3%), while models with Mn ($m_{ab}$, $m_c$) and Nd ($m_{ab}$) or Mn ($m_{ab}$, $m_c$) yielded an improved fit ($R_{wp}$ = 5.1%). This suggests that there is an intermediate phase with a Mn moment rotation angle (φ) in between 0-90° with respect to the $c$-axis. However, the comparable fits also demonstrate that this phase exists only in a narrow temperature interval, which is in agreement with the heat capacity and magnetic susceptibility data. The temperature dependence of the ordered moments [using the model with Mn ($m_{ab}$, $m_z$) and Nd ($m_{ab}$) between $20 < T < 25$ K] are given in Fig. 6c. The fitted $\phi$ values are 12(10)° at 23.6(2) K, 29(5)° at 21.7(2) K, and 80(9)° at 19.9(2) K. This suggests that the SR temperature interval is $< 3$ K. The temperature evolution of the ordered Nd moment is of some interest: an initial rapid increase to about 0.9 $\mu_B$ consistent with a 2$^{nd}$ order phase transition is observed but is followed by a gradual increase in moment to a maximum value of 1.9 $\mu_B$ at 4 K, which corresponds to the 2$^{nd}$ transition observed in the magnetic susceptibility. There is no evidence for a phase transition to a different magnetic structure at 4 K as the statistics above and below are comparable, and no significant changes in magnetic intensities are observed. The temperature evolution of the ordered Mn moment above 50 K is shown in the inset to Fig. 6c. A fit to $m = m_0(1-T/T_N)^\beta$ yields $m_0 = 3.85(2)$ $\mu_B$ K, $T_N = 359(2)$ K, and $\beta = 0.27(1)$.

**Discussion**

The magnetic and electronic properties of NdMnAsO are of much interest, in particular in comparison with the isostructural Fe, Co and Ni materials, which are all itinerant systems. From a structural point of view the most pronounced difference is the unit cell volume (or the MnAs bond distance), which is much larger than expected based on a simple extrapolation of the itinerant systems. This signals significant differences in bonding, which is confirmed by the local moment magnetism and semiconducting behavior observed in the title material. The main novel result comes from the neutron powder diffraction study that reveals a significant coupling between the Nd and Mn magnetic sublattices, which results in a spin-reorientation of the Mn moments. This type of transition has not been observed in any of the itinerant 1111-type systems, although it has been postulated for PrFeAsO.[20] The analysis of the neutron powder diffraction data demonstrates that the SR of the Mn moments occurs in a narrow temperature interval ($< 3$K), and proceeds via an intermediate phase with a Mn rotation angle between 0-90° with respect to the *c*-axis. This interpretation is in agreement with a recent pre-print on PrMnSbO that reveals an identical SR transition occurring over a much wider (~10 K) temperature interval.[32] The narrow SR transition for NdMnAsO is consistent with the magnetic susceptibility and heat capacity measurements that evidence a 2$^{nd}$ order phase transition at $T_{SR}$. The probable cause of the SR is the competing single ion anisotropy of Nd and Mn as observed in the RFeO$_3$ perovskites,[33-35] and the high-$T_c$ parent materials R$_2$CuO$_4$,[36] which have all been accounted for using a phenomenological description involving competing single ion anisotropies. In the current case, the Mn$^{2+}$ ion (free ion values: S = 5/2, orbital angular momentum L = 0) prefers to orient its moment along the *c*-axis, while the Nd$^{3+}$ moment (free ion values: S = 3/2 and L = 6) lies in the basal plane. Since the single ion anisotropy of Nd$^{3+}$ dominates that of Mn$^{2+}$ (L = 0) a SR transition of the Mn moments can occur as soon as an ordered moment develops on the Nd sublattice. The preferred basal plane orientation for the Nd moments is supported by calculations of the crystalline electric field splitting of the 4f states.[36] The development of a small magnetization on the Nd sublattice (Fig. 6c), followed by an almost linear increase to the saturation value has also been observed in a recent single crystal study on NdFeAsO,[37] and it may be that this is a general feature of these materials. Further theoretical analysis is needed to explain this unexpected behavior, and to elucidate the magnetic coupling between the Nd and TM sublattices. At RT only the Mn sublattice is magnetically ordered and has a NN AF checkerboard alignment in the basal plane (Fig. 6a). The coupling between Mn planes is FM, and the ordered moment is aligned along the *c*-axis. The in-plane checkerboard arrangement suggests that next nearest neighbor (NNN) interactions are negligible. This is in contrast to NdFeAsO, where in a localized model strong NNN interactions are needed to stabilize the observed spin-stripe ordering.[38] Within the insulating Nd-O$_2$-Nd blocks the magnetic order is characterized by AF coupled FM planes (Fig. 6b). An identical arrangement occurs for NdCoAsO that also remains tetragonal to the lowest measured temperatures.[14, 15] In contrast, the Cmma ($\sqrt{2}a \times \sqrt{2}a \times c$) superstructure for NdFeAsO ($\mathbf{k}$ = 0) allows for a different type of ordering with AF coupled AF Nd planes.[38] To further analyze the observed SR a schematic representation of the main magnetic interactions in NdMnAsO is presented in Fig. 7. Within the unit cell, the magnetic ions form a diamond shape with Nd at the top and bottom corners. As discussed, the AF NN Mn-Mn interaction $J_1$ (through-space distance d = 2.86 Å) is dominant and leads to the observed checkerboard magnetic ordering. Within the AF Nd-O$_2$-Nd blocks there are two main interactions: $J_3$ which cuts across the block (Nd1-Nd2, d = 3.68 Å), and $J_4$ which is along the block (Nd$n$-Nd$n$, n = 1,2, d = 4.04 Å). Based on the Nd-O-Nd bond angles (Table 1) and the Kanamori-Goodenough rules for superexchange both are expected to be AF, although only $J_3$ is in the experimental structure. The interaction between the Nd and Mn sublattices proceeds via four identical paths ($J_2$, d = 3.86 Å), which leads to a frustrated arrangement of symmetric Heisenberg-exchange interactions between local moments. This leaves the weaker antisymmetric (Dzyaloshinsky-Moriya) exchange as the strongest magnetic interaction, and a postulated perpendicular orientation of the ordered Mn and Nd moments. Unfortunately, due to the tetragonal symmetry this cannot be verified experimentally. Finally, the Neel temperature (359(2) K) is much reduced compared to the 625 K reported for BaMn$_2$As$_2$.[19] In addition, the critical exponent [$\beta$ = 0.27(1)] is lower than reported for BaMn$_2$As$_2$ [$\beta$ = 0.35(2)], which is close to the expected value for a 3-dimensional Heisenberg magnet ($\beta$ = 0.367).[19] These observations point to a reduced magnetic coupling and more 2-dimensional behavior in this 1111-type material compared to the 122's, which is plausible given the larger distance between the Mn layers. The extrapolated ordered moment at 0 K is 3.85(2) $\mu_B$, which is identical to the value reported for BaMn$_2$As$_2$ (3.88(4) $\mu_B$ at 10 K).[19]

To conclude: NdMnAsO is a local moment semiconductor that exhibits a Mn spin-reorientation transition from aligned along the c-direction into the basal plane upon ordering of the Nd moments.

**Acknowledgements**

EPSRC-GB is acknowledged for the provision of beam time at the ILL and ESRF.

**Figure Captions**

Fig. 1. (color online) Rietveld fit to 4 K synchrotron X-ray powder diffraction data for NdMnAsO. Observed data are indicated by open circles, the fit by the solid line, and the difference curve is shown at the bottom. The Bragg markers are for NdMnAsO (top), MnAs (middle) and $Nd_2O_3$ (bottom).

Fig. 2. (color online) (a) Temperature dependence of the crystallographic *a*- and c-axes for NdMnAsO. (b) Temperature dependence of the unit cell volume and the c/a ratio for NdMnAsO.

Fig. 3. (color online) Temperature dependence of the ZFC magnetic ($\chi$) and inverse magnetic ($1/\chi$) susceptibilities for NdMnAsO. The solid line is a fit to the Curie-Weiss law. The insets show the field dependence of the magnetization at 2, 15 and 30 K, and a close up of $\chi(T)$ at the Nd ordering transition.

Fig. 4. (color online) Temperature dependence of the electrical resistivity for NdMnAsO. The inset shows the field dependence of the normalized resistivity $R/R_0$ at 2, 15, 30, 100 and 250 K.

Fig. 5. (color online). Temperature dependence of the heat capacity (C) in applied magnetic fields of 0 and 9 Tesla. The inset shows a linear fit to C/T versus $T^2$ at low temperatures.

Fig. 6. (color online) Rietveld fits to 30 K (a) and 1.6 K (b) neutron powder diffraction data for NdMnAsO. The observed data are indicated by open circles, the fit by the solid line, and the difference curve is shown at the bottom. From top to bottom: Bragg markers are for NdMnAsO, the magnetic phase, $Nd_2O_3$ (2.0(2) wt%) and MnAs (1.0(2) wt%). (30 K: $R_{wp}$ = 5.1%, $R_p$ = 3.2% and $R_F^2$ = 3.4%, 1.6 K: $R_{wp}$ = 5.5%, $R_p$ = 3.5% and $R_F^2$ = 5.1%). The insets to (a) and (b) show graphic representations of the fitted magnetic structures. The temperature evolution of the fitted Nd and Mn moments between 1.6 and 50 K is given in (c). The inset shows the temperature

evolution of the ordered Mn moment between 50-300 K. The solid line is a fit to $m = m_0(1-T/T_N)^\beta$ (see text). The unidentified reflections denoted by the asterisks do not change intensity between 1.6 and 400 K.

Fig. 7. (color online) Schematic representation of the main magnetic interactions for NdMnAsO. Within the unit cell, the Nd and Mn ions lie on a diamond shaped plane. The AF NN interactions ($J_1$) between Mn ions are all satisfied, leading to G-type magnetic ordering. The symmetric Heisenberg exchange interaction ($J_2$) between the Nd and Mn sublattice is frustrated, leaving antisymmetric (Dzyaloshinsky-Moriya) exchange as the strongest magnetic interaction, and a postulated perpendicular orientation of the ordered Mn and Nd moments. Within the Nd-$O_2$-Nd blocks the strongest magnetic interaction ($J_3$) is antiferromagnetic, while $J_4$ is ferromagnetic. Magnetic interactions between the As-$Mn_2$-As (Nd-$O_2$-Nd) blocks are expected to be much weaker due to the large separation between the blocks. The relative orientations of the ordered magnetic moments within each sublattice at 1.6 K are denoted by the +/- signs. The labeling of the atoms corresponds to that used in Table II.

Table I. Refined lattice constants, atomic parameters, selected bond lengths (Å) and angles (°), and fit statistics for NdMnAsO from Rietveld fits against synchrotron X-ray powder diffraction data.

| T (K) | | 4 | 100 | 250 |
|---|---|---|---|---|
| SG | | P4/nmm | P4/nmm | P4/nmm |
| a-axis (Å) | | 4.04446(1) | 4.04503(1) | 4.04979(5) |
| c-axis (Å) | | 8.87383(2) | 8.88258(2) | 8.89935(2) |
| Volume (Å$^3$) | | 145.16 | 145.34 | 145.97 |
| Nd | $U_{iso}$(Å$^2$) | 0.0010(1) | 0.0025(1) | 0.0053(1) |
| | z | 0.12991(5) | 0.12992(4) | 0.12976(4) |
| | frac | 1.00 | 1.00 | 1.00 |
| Mn | $U_{iso}$(Å$^2$) | 0.0017(1) | 0.0039(2) | 0.0073(2) |
| | frac | 1.00 | 1.00 | 1.00 |
| As | $U_{iso}$(Å$^2$) | 0.0013(1) | 0.0044(2) | 0.0071(2) |
| | z | 0.6737(1) | 0.6736(1) | 0.6736(1) |
| | frac | 1.00 | 1.00 | 1.00 |
| O | $U_{iso}$(Å$^2$) | 0.0041(1) | 0.0046(1) | 0.009(1) |
| | frac | 1.00 | 1.00 | 1.00 |
| d(Nd-O) (Å) | | 2.3277(2) | 2.3286(2) | 2.3308(2) |
| ∠Nd-O-Nd (°) 2x | | 120.63(2) | 120.58(2) | 120.63(2) |
| ∠Nd-O-Nd (°) 4x | | 104.22(2) | 104.22(2) | 104.21(2) |
| d(Mn-As) (Å) | | 2.5456(5) | 2.5424(4) | 2.5468(4) |
| ∠As-Mn-As (°) 2x | | 111.65(2) | 111.54(1) | 111.59(2) |
| ∠As-Mn-As (°) 4x | | 105.20(3) | 105.41(2) | 105.32(2) |
| d(Mn-Mn) (Å) | | 2.85983(1) | 2.86027(1) | 2.86364(1) |
| $\chi^2$ | | 5.1 | 4.7 | 5.1 |
| $wR_p$ | | 12.8 | 14.1 | 15.0 |
| $R_p$ | | 8.3 | 9.6 | 9.6 |
| $R_F^2$ | | 4.3 | 5.5 | 5.1 |

Nd: 2c (¼, ¼, z); Mn: 2b (¾, ¼, ½); As: 2c (¼, ¼, z); O: 2a (¾, ¼, 0)

Impurities: 1.5(1) wt% $Nd_2O_3$, 1.0(1) wt% MnAs.

Table II. Basis vectors [$m_x$, $m_y$, $m_z$] for space group P4/nmm with **k** = 0, and $R_{wp}$ values for fits against the 1.6 K and 30 K datasets.

|  | $\Gamma_2^1$ | $\Gamma_3^1$ | $\Gamma_6^1$ | $\Gamma_9^2$ | $\Gamma_{10}^2$ |
|---|---|---|---|---|---|
| Mn1 | - | [0 0 $m_z$] | [0 0 $m_z$] | [$m_x$ $m_y$ 0] | [$m_x$ $m_y$ 0] |
| Mn2 | - | [0 0 $m_z$] | [0 0 -$m_z$] | [$m_x$ $m_y$ 0] | [-$m_x$ -$m_y$ 0] |
| Nd1 | [0 0 $m_z$] | [0 0 $m_z$] | - | [$m_x$ $m_y$ 0] | [$m_x$ $m_y$ 0] |
| Nd2 | [0 0 -$m_z$] | [0 0 $m_z$] | - | [$m_x$ $m_y$ 0] | [-$m_x$ -$m_y$ 0] |
| $R_{wp}$, 30 K (%) | unstable | - | 5.1 | - | 16.3 |
| $R_{wp}$, 1.6 K (%) | 10.5 | - | 9.3 | - | 5.4 |

Mn1: (0.75, 0.25, 0.5), Mn2: (0.25, 0.75, 0.5), Nd1: (0.25, 0.25, 0.139) and Nd2: (0.75, 0.75, 0.861).

Fig. 1

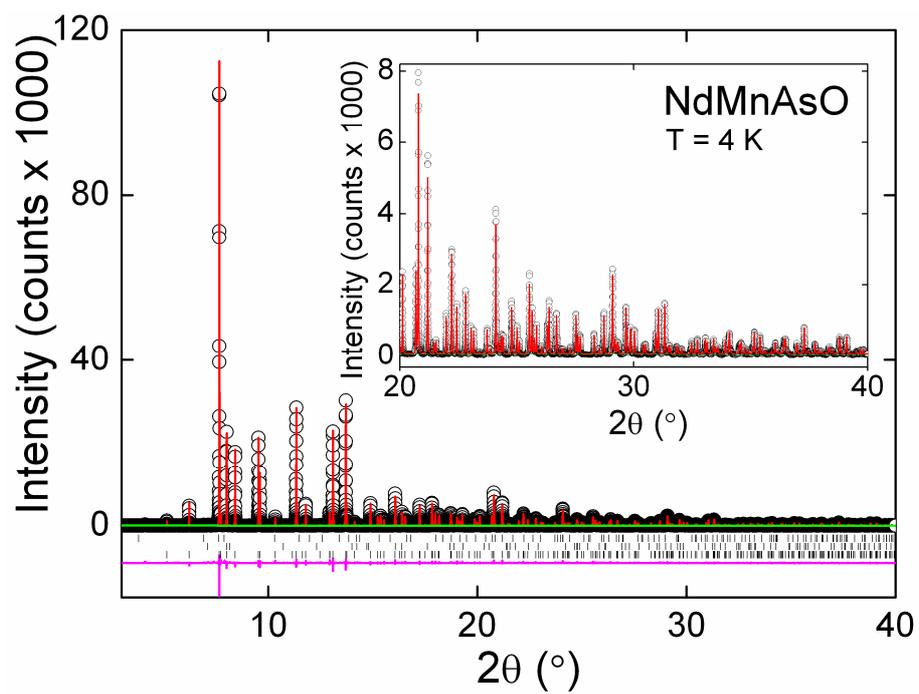

Fig. 2a

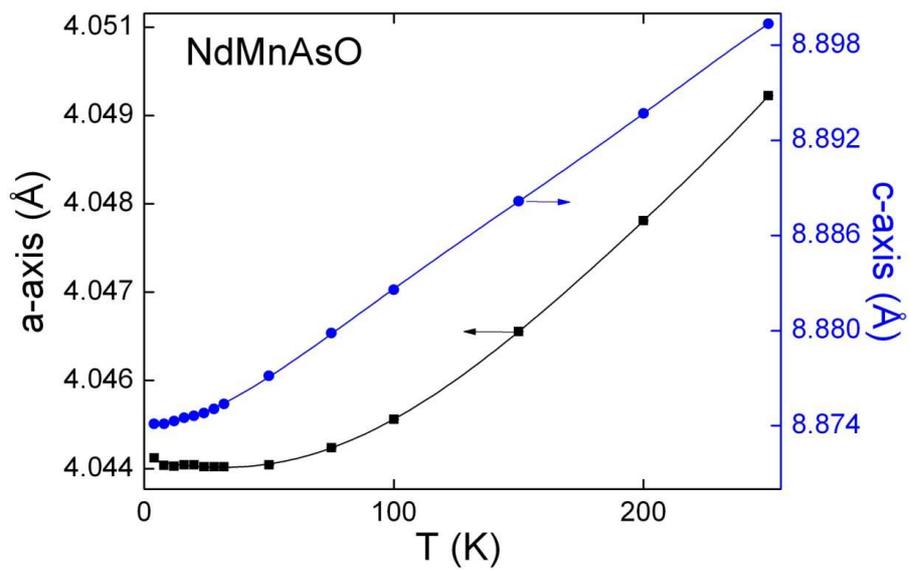

Fig. 2b

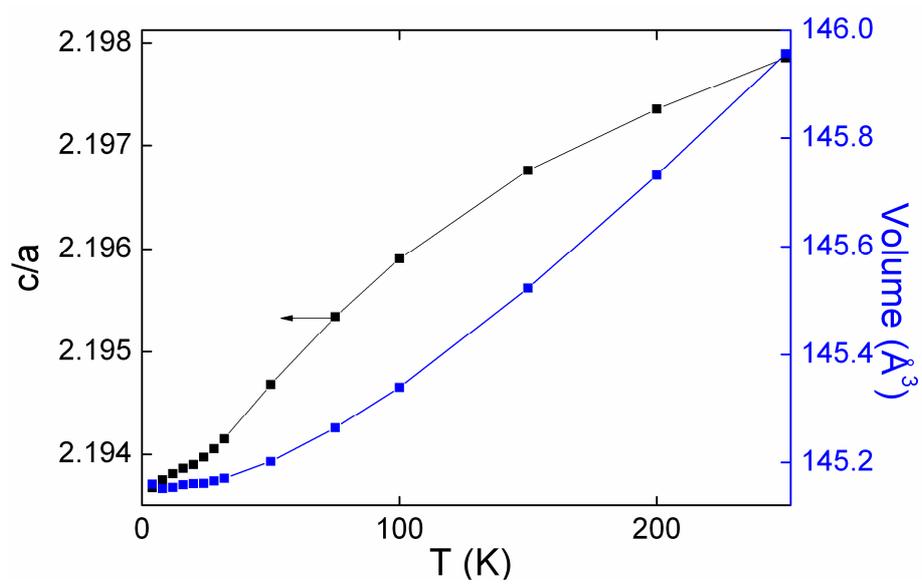

Fig. 3

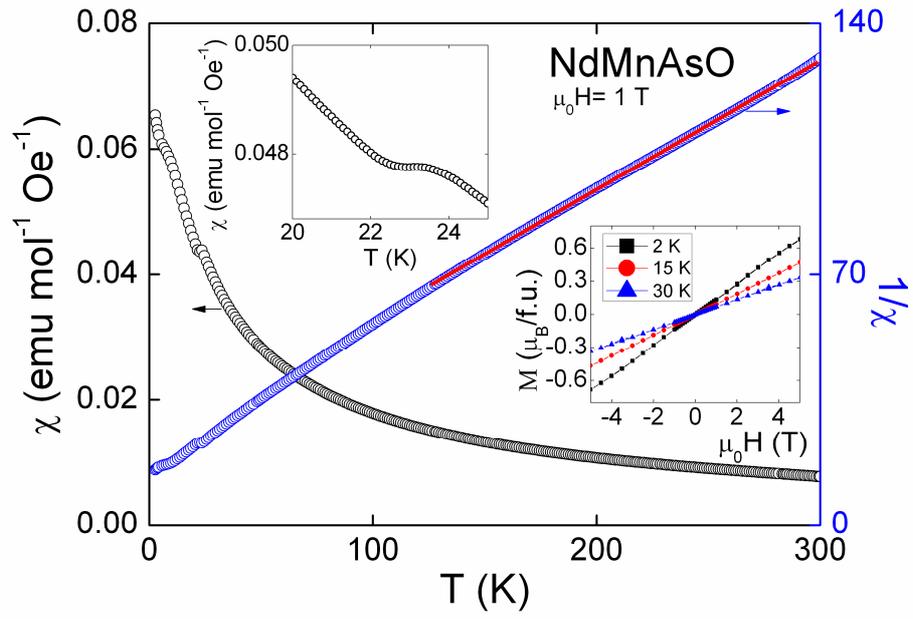

Fig. 4

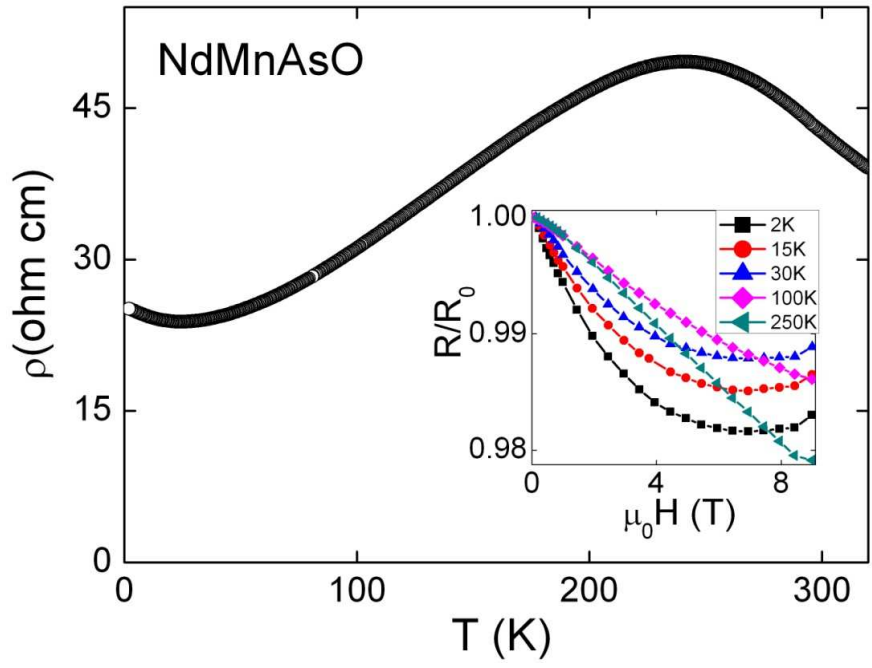

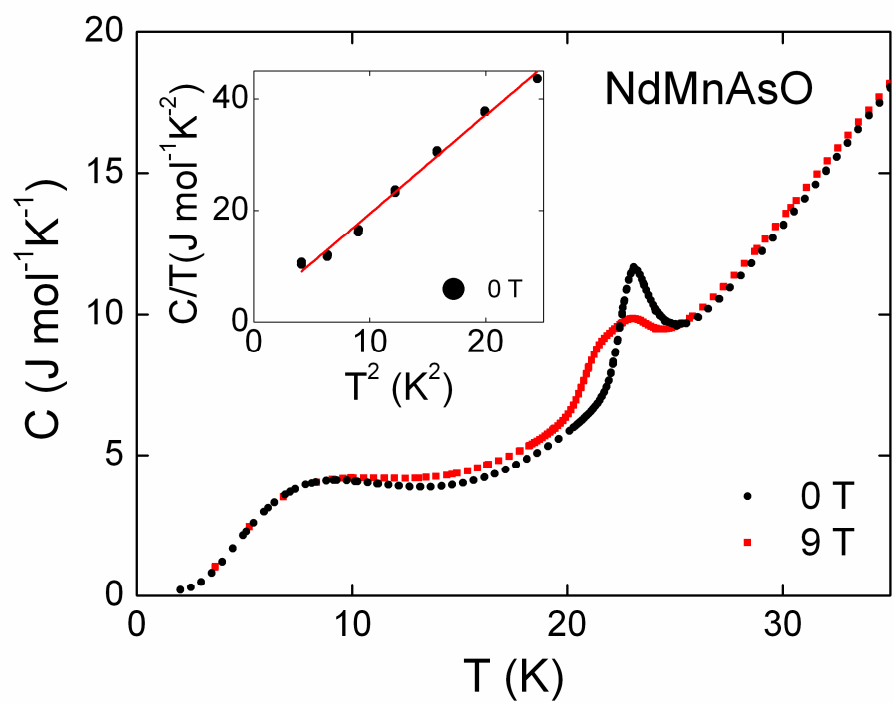

Fig. 6a

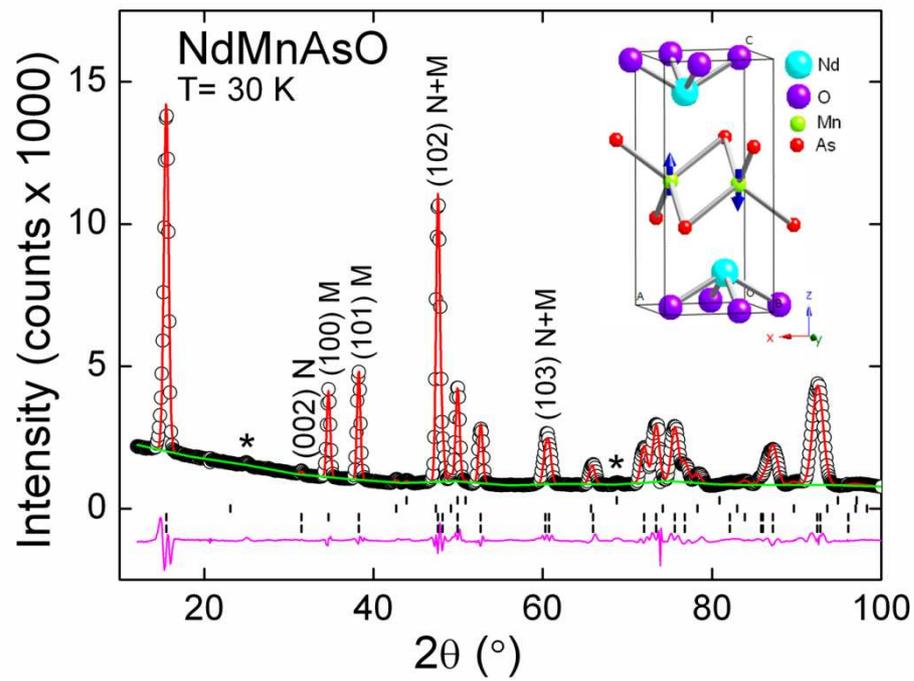

Fig. 6b

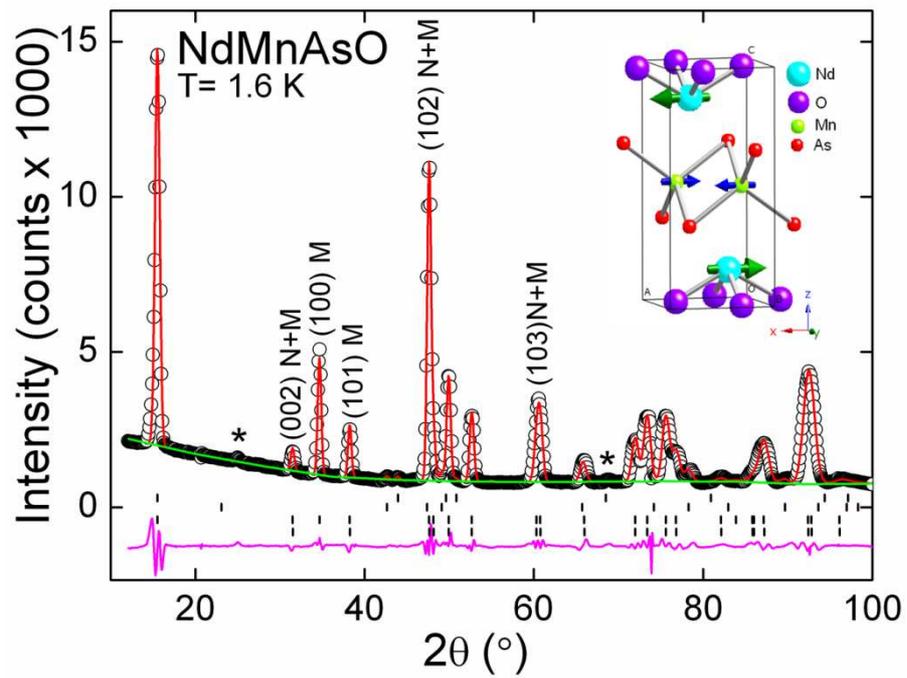

Fig. 6c

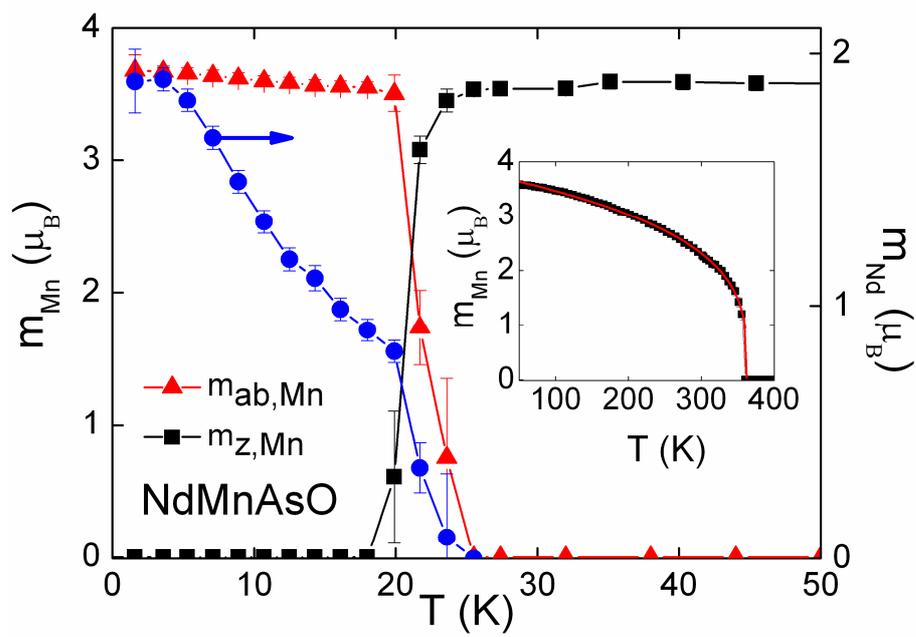

Fig. 7.

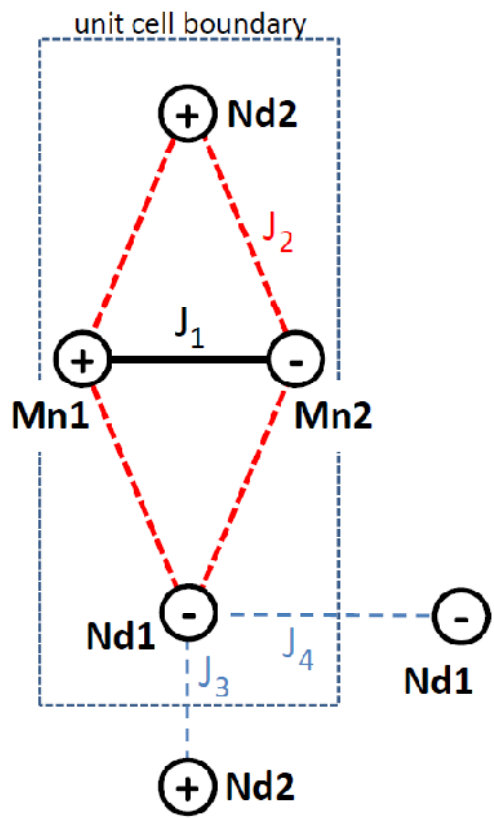